\documentclass[sigconf]{acmart} % to see pdf version for PUBLICATION-ready
\AtBeginDocument{%
  }

\setcopyright{acmlicensed}
\copyrightyear{2026}
\acmYear{2026}
\acmConference[DIS Companion '26]{Designing Interactive Systems Conference}{June 13--17, 2026}{Singapore, Singapore}

\usepackage{graphicx}
\usepackage{url}   % Ensures URLs break properly
\usepackage{xcolor}
\usepackage{soul}
\usepackage[most]{tcolorbox}
\usepackage[dvipsnames]{xcolor}
\usepackage{subfiles}
\usepackage{makecell}
\usepackage{enumitem}
\usepackage{multirow}
\usepackage{longtable}
\usepackage{graphicx}
\usepackage{subcaption}
\usepackage{listings}
\usepackage{pgfplots}
\usepackage{pgf-pie}
\usepackage{soul}
\pgfplotsset{compat=1.17}
\usepackage{amsmath}
\usepackage{array}
\usepackage{float}
\usepackage{pdflscape}
\usepackage{rotating}
\usepackage{csquotes}
\usepackage{CJKutf8}

\newif\ifredact
\redacttrue   % Uncomment this line for the redacted version
\newif\ifcomment
\commenttrue   % Comment this line to hide comments
\ifcomment
    \newcommand{\missing}[2][]{\textcolor{red}{[\textbf{MISSING\ifx#1\empty\else~–~#1\fi}] ~#2}}
    \newcommand{\kel}[1]{~\sethlcolor{pink}\hl{[Kellie: #1]}}
    \newcommand{\ken}[1]{~\sethlcolor{yellow}\hl{[Kenny: #1]}}
\else
  \newcommand{\missing}[2]{}
  \newcommand{\kel}[1]{}
  \newcommand{\ken}[1]{}
\fi

\newcommand{\earnprom}{\textbf{\textit{PS1}}}
\newcommand{\bluemoon}{\textbf{\textit{PS2}}}
\newcommand{\warmsun}{\textbf{\textit{PS3}}}
\newcommand{\jaderiver}{\textbf{\textit{PS4}}}
\newcommand{\windsong}{\textbf{\textit{PS5}}}
\newcommand{\silvercove}{\textbf{\textit{PS6}}}
\newcommand{\strongarm}{\textbf{\textit{PS7}}}
\newcommand{\clearsky}{\textbf{\textit{PS8}}}
\newcommand{\calmsea}{\textbf{\textit{PS9}}}
\newcommand{\ironpetal}{\textbf{\textit{PS10}}}
\newcommand{\echoblaze}{\textbf{\textit{PS11}}}
\newcommand{\highpeak}{\textbf{\textit{PS12}}}
\newcommand{\shadowdancer}{\textbf{\textit{PS13}}}
\newcommand{\redbird}{\textbf{\textit{PS14}}}
\newcommand{\duskytide}{\textbf{\textit{PS15}}}
\newcommand{\brightstar}{\textbf{\textit{PS16}}}
\newcommand{\crystalstream}{\textbf{\textit{PS17}}}
\newcommand{\bluepea}{\textbf{\textit{PS18}}}
\newcommand{\flowymoon}{\textbf{\textit{PS19}}}
\newcommand{\butterflywave}{\textbf{\textit{PS20}}}

\newcommand{\dquote}[1]{\enquote{#1}}
\newcommand{\squote}[1]{\enquote*{#1}}

\begin{document}
\renewcommand\footnotetextcopyrightpermission[1]{} % Removes copyright footnote
\settopmatter{printacmref=false} % Removes citation information below abstract
% \title{Towards Human-Centred AI for Peer Support}\ken{a bit too high level, prob needs a bit more detailed a title}
\title[Emotional Labour, Responsibility, and AI in Peer Support]{"I'm Not Able to Be There for You": Emotional Labour, Responsibility, and AI in Peer Support}

\author{Kellie Yu Hui Sim}
\email{kellie_sim@mymail.sutd.edu.sg}
\orcid{0009-0005-6451-7089}
\affiliation{
  \institution{Singapore University of Technology and Design}
  \city{Singapore}
  \country{Singapore}
}
\authornote{Corresponding author. This is a preprint of the paper accepted at DIS 2026. The final version will be available in the ACM Digital Library.}

\author{Kenny Tsu Wei Choo}
\email{kenny_choo@sutd.edu.sg}
\orcid{0000-0003-3845-9143}
\affiliation{
  \institution{Singapore University of Technology and Design}
  \country{Singapore}
}

\renewcommand{\shortauthors}{Sim and Choo}

\begin{abstract}
    % fewer than 150 words 
% v1
% Peer support is increasingly positioned as a scalable response to gaps in mental health care, particularly in digitally mediated settings.
% However, what constitutes peer support, who counts as a peer supporter, and how responsibility is distributed remain unevenly defined in practice.
% In this paper, we examine how peer support is enacted within the Singapore context, where heterogeneous entry pathways, informal caregiving, and institutional expectations coexist.\ken{I'm not sure stating Singapore here is needed--the value is more on the last part of this sentence on entry pathways etc..}
% We interviewed peer supporters, we surface how lived experience, moral commitment, and self-identification shape participation in peer support, while also blurring boundaries around scope, authority, and accountability.
% Participants described substantial emotional and cognitive labour, with responsibility for boundary-setting, escalation, and self-care frequentlycon absorbed at the individual level in the absence of consistent organisational support.
% Their evaluations of AI further reveal that support technologies are judged less by technical capability than by how they redistribute risk, labour, and responsibility.

% v2
Peer support is increasingly positioned as a scalable response to gaps in mental health care, particularly in digitally mediated settings, yet what counts as peer support and how responsibility is distributed remain unevenly defined in practice.
Drawing on interviews with peer supporters, we show how lived experience, moral commitment, and self-identification shape participation while blurring expectations around scope, authority, and accountability.
Institutional ambiguity concentrates emotional labour, boundary-setting, and escalation of responsibility at the individual level, often without consistent organisational scaffolding.
Participants evaluated AI not primarily through empathy or technical capability, but through how technologies redistribute risk, labour, and accountability within already fragile support roles.
Building on these findings, we outline design futures for an AI-supported peer support ecosystem that foregrounds responsibility as a central design concern rather than treating AI as a mechanism of scale.
\end{abstract}

\begin{CCSXML}
<ccs2012>
   <concept>
       <concept_desc>Human-centered computing~Empirical studies in HCI</concept_desc>
       </concept>
 </ccs2012>
\end{CCSXML}

\ccsdesc[500]{Human-centered computing~Empirical studies in HCI}

% \keywords{Do, Not, Use, This, Code, Put, the, Correct, Terms, for,
\keywords{Peer Support, Mental Health, Volunteer, Digital Peer Support, AI, Human-Centred AI}

% \received{13 February 2026}

\maketitle

\section{Introduction and Related Work}
Peer support is increasingly positioned as a scalable response to gaps in mental health care, particularly in digitally mediated and community-based settings.
Often framed as relational, accessible, and grounded in mutual lived experience~\cite{yeoDigitalPeerSupport2023, shalabyPeerSupportMental2020, solomonPeerSupportPeer2004}, peer support is treated in much of the literature as a relatively coherent practice with shared norms around training, scope, and responsibility~\cite{leePeerSupportMental2019, frankeImplementingMentalHealth2010}.
At the same time, emerging digital platforms and AI systems are proposed as mechanisms to extend capacity, scaffold conversations, and standardise quality~\cite{hsuHelpingHelperSupporting2025, sharmaHumanAICollaboration2023}.
In this paper, we focus on AI systems that directly mediate or scaffold peer support interactions, particularly large language model (LLM)-based systems.
We refer to these collectively as AI-supported peer support infrastructures.

Yet what constitutes \emph{peer support}, who counts as a peer supporter, and how responsibility is distributed remain unevenly defined in practice.
We refer to this coexistence of differing expectations as \emph{definitional plurality}.
In practice, this requires supporters to negotiate boundaries and expectations in situ, often without clear guidance on scope or responsibility.
Prior work has documented volunteer burnout, turnover, and training gaps~\cite{yaoLearningBecomeVolunteer2022, chenScaffoldingOnlinePeersupport2021}, but these are often treated as operational constraints rather than as consequences of ambiguous authority.
Research on emotional labour further suggests that care work is sustained through affective regulation, boundary negotiation, and responsibility management~\cite{poonComputerMediatedPeerSupport2021}, yet this lens is rarely centred in the design of peer support technologies.

This tension is particularly visible in Singapore, where peer support spans school-based programmes, community initiatives, and digital platforms~\cite{yeoDigitalPeerSupport2023, simSaidThingsNeeded2025}.
While training may be provided to supporters, individuals may self-identify as peer supporters regardless of whether they have lived experience or professional credentials.
At the same time, institutions increasingly rely on them to fill systemic gaps in mental health provision.
The label "peer supporter" therefore carries strong moral and interactional expectations while remaining ambiguously defined.

As a result, definitional plurality shapes how peer support is enacted in practice.
Peer supporters describe negotiating boundaries around availability, escalation, and emotional involvement, particularly in text-based settings where cues are limited~\cite{iftikharTogetherNotTogether2023}.
When expectations exceed capacity, responsibility is frequently absorbed at the individual level rather than redistributed through organisational or technical infrastructures~\cite{kimSupportersFirstUnderstanding2023}.

Meanwhile, AI systems designed to scaffold peer support~\cite{hsuHelpingHelperSupporting2025, steenstraScaffoldingEmpathyTraining2025} implicitly assume particular models of authority, training, and accountability.
When these assumptions fail to align with local understandings, technologies risk amplifying emotional load, obscuring responsibility, or inflating perceived authority without corresponding support~\cite{bhattacharjeeInvestigatingRoleSituational2025, songTypingCureExperiences2025}.

In this paper, we examine how divergent definitions of peer support shape lived practice, emotional labour, and boundary-setting, and consider the implications for designing responsibility-aware technologies.
Drawing on interviews with peer supporters, we surface how participants describe their roles, challenges, and expectations around AI.
We argue that accounting for definitional plurality is essential for designing context-sensitive systems that support peer supporters without reproducing hidden burdens or misaligned assumptions about responsibility and scale.
We also use these findings to articulate design futures for AI-supported peer support, and propose a responsibility-aware ecosystem that redistributes labour more equitably, safeguards supporter well-being, and preserves human judgement.
\section{Methodology}
% We recruited 20 trained peer supporters (14 female) aged 18–54 ($M=29.45$, $SD=8.17$) with prior training in peer support and/or Psychological First Aid~\cite{shultzPsychologicalFirstAid2014a}.
% Participants reported peer support experience ranging from 6 months to over 10 years ($M=4.28$ years) and were engaged in online-only ($n=3$), offline-only ($n=6$), or hybrid ($n=11$) support settings within community-based and institutionally affiliated peer support programmes in Singapore.
We interviewed 20 trained peer supporters in Singapore (14 female; aged 18-54) with 6 months to over 10 years of experience across online, offline, and hybrid settings.
Participants came from diverse academic and professional backgrounds, with varying levels of formal psychology training.
All participants received training through their respective organisations, typically in peer support and/or Psychological First Aid~\cite{shultzPsychologicalFirstAid2014a}. 
A detailed breakdown of participant demographics and experience is provided in Appendix~\ref{appendix:demographics}.
During the study, participants completed a demographic questionnaire, followed by a semi-structured interview ($M\simeq38$ min, $SD\simeq11$ min) that covered pathways to peer support, emotionally challenging experiences, boundary management, and perspectives on digital or AI-supported mental health tools (see Appendix~\ref{appendix:interview-guide}).
%Interviews lasted between 25 minutes and 19 seconds and 1 hour, 14 minutes and 57 seconds ($M = 38{:}22$, $SD = 11{:}07$).
Interviews were audio-recorded, transcribed using WhisperX~\cite{bain2022whisperx}, and manually corrected.
The study received Institutional Review Board approval (SUTD IRB-24-00670).
All procedures followed ethical research guidelines.
Participants received a $\sim$USD\$7.70 voucher as compensation.

Using framework analysis~\cite{galeUsingFrameworkMethod2013} combining deductive and inductive coding, three researchers independently coded transcripts.
Codes were iteratively refined through discussion, with disagreements resolved through consensus.
The agreed-upon codes were then applied to the full dataset, and the resulting codes were compared across participants and iteratively grouped into higher-level themes.
\section{Results}
% - lived experiences [brief??? to show that not all are firsthand] (4.1.2)
% - skills, frameworks, gaps (4.2)
% - how people wanted tech / AI to be used (4.3.3)

\subsection{Lived Experience as Catalyst: Reciprocity and Moral Commitment in Peer Support}
% Who becomes a peer supporter, and why this complicates definitions
Participants entered peer support through both firsthand and secondhand exposure to distress, often becoming informal \dquote{go-to} figures before assuming structured roles.
These were often informal pathways, complicating definitions and boundaries of peer support.
While these experiences grounded empathy and positioned peer support as a reciprocal practice (rather than purely voluntary), they also heightened awareness of emotional cost and limits of responsibility.
Many described the need for boundaries, self-regulation, and ongoing reflection to avoid exhaustion or burnout.

13 participants described firsthand experiences with burnout, mental health conditions such as anxiety or depression, or suicidal ideation.
For some, recovery from these experiences motivated a desire to support others who were struggling.
As \echoblaze{} reflected, \dquote{I had very heavy mental issues but then I [...] came to a place of mental capacity that I can help others [...] I'm very grateful of the therapy I received that I want to give back}. 

Others entered peer support through secondhand exposure, including family members with chronic depression, friends in suicidal crises, or sustained struggles in various settings.
These experiences shaped participants' understanding of the limits of informal care and heightened their sense of responsibility toward those in distress.
As \warmsun{} shared, \dquote{I feel the sadness coming to me as well [...] That was my challenge because of my family history [...] I've learned how to step backwards and manage both their emotions and our emotions}.

\subsection{Situated Practice and Emotional Demands in Peer Support}
% How peer support is practised, and where responsibility falls
Across school, community, and online contexts, participants described peer support as situated and emotionally demanding.
Moments of perceived impact and affirmation were described by many participants, including instances in which peer supporters observed shifts in the care recipient's emotional state or engagement (e.g., reduced distress, continued disclosure) or received explicit or implicit validation of their support.
However, these experiences were consistently interwoven with emotional strain, boundary negotiation, and uncertainty around role expectations.

Participants practised peer support across diverse formal and informal contexts.
\bluemoon{} scheduled calls or in-person meetings when necessary, while \duskytide{} sustained anonymous long-term one-to-one relationships online.
Yet the definition of \squote{peer support} itself was not always clear-cut.
As \jaderiver{} observed, mixed perceptions and the industry's complex state complicated expectations regarding legitimacy, scope, and responsibility.

Emotional and cognitive labour were consistently emphasised, but at the level of individual peer supporters.
Participants described anxiety about misjudging readiness, saying the wrong thing, or failing to provide containment in text-based interactions.
\bluemoon{} noted the need to \dquote{soothe [the care recipient's] emotions first}, describing the need to process words three to four times as draining.
Others described burnout and emotional depletion after prolonged engagement, especially when empathy was sustained without clear boundaries.
As \silvercove{} noted, \dquote{If you empathise too much [...] you're not really taking care of yourself}.

Participants also described persistent role ambiguity, particularly when expected to support peers beyond their capacity.
Balancing confidentiality with safety, interpreting emotional cues, and deciding when to escalate were often guided by personal judgement rather than formal protocol.
To cope, participants developed self-devised strategies such as boundary-setting scripts, modelling counsellors, or temporary disengagement.
For example, \silvercove{} shared, \dquote{I would tell them, \squote{I'm sorry, I'm not able to be there for you [...] because I'm going through something on my own.}}.
Others described using non-confrontational techniques to avoid invalidating peers while maintaining emotional distance, such as by not disagreeing with peers and avoiding jumping to conclusions.
However, boundary management was largely treated as an individual responsibility rather than an institutional one.
Expectations around availability, emotional containment, and continuity of care were often implicit rather than institutionally reinforced.
As \windsong{} reflected, peers sometimes expected constant presence or solutions, even when none were possible.

Organisational support varied considerably, resulting in uneven distributions of responsibility across settings.
While some organisations had frameworks such as escalation protocols, phrasing guides, or supervision structures, these supports varied widely across settings.
More structured programmes provided greater confidence in managing scope and responsibility, such as through colour-coded check-ins that allowed supporters to step back without stigma (\echoblaze{}).
Others noted minimal guidance beyond generic reminders to \dquote{do your self-care} (\calmsea{}), leaving emotional sustainability largely unaddressed.
Those without guidance relied on self-developed practices (e.g., boundary-setting, rest, emotional withdrawal) largely learnt through burnout or emotional depletion.
Several emphasised the need to disengage when emotionally depleted.
For example, \brightstar{} noted: \dquote{The best self-care was taking a short break [...] do anything but peer support}.
These strategies reflected an awareness of personal limits, but also highlighted the absence of shared mechanisms for redistributing care labour.
The heavy reliance on individual self-regulation amid uneven infrastructural support shaped engagement and, in some cases, led to withdrawal.

\subsection{Integrating AI into Peer Support: Constraints, Conditions, and Concerns}
% What kinds of technology peer supporters accept or resist, and why
Participants expressed ambivalence about integrating AI into peer support.
While AI was sometimes seen as a useful aid, participants consistently resisted framings that positioned AI as a replacement for human support.
Instead, they evaluated AI in terms of risk, responsibility, and its potential impact on emotional labour and boundary-setting.

Several participants suggested that AI could play a limited role in low-stakes situations, such as when peers were experiencing mild distress or required basic emotional scaffolding.
In these contexts, AI was viewed as a temporary support that could maintain engagement or structure conversations until human support became available.
This would assist interaction without assuming responsibility for outcomes.
As \bluemoon{} explained: \dquote{Let me define it in a crisis level [...] One [...] is a situation that happened but it's not affecting the person a lot [...] Ten is when the person is [...] about to crash [...] If the person is between one to three [...] AI might help. But if it's three and above [...] they'll see the letters but won't be reading.}
Across accounts, participants emphasised that such use required clear thresholds and seamless escalation pathways, with AI remaining subordinate to human judgement.

Some participants strongly rejected the idea of AI substituting for human peer supporters in emotionally complex or high-risk interactions.
Rather than delegating responsibility to AI, they envisioned human-AI collaboration in which AI supports, but does not direct, peer support work.
As \crystalstream{} noted, \dquote{The human touch still needs to be explored [...] maybe a human face behind the AI once in a while}.
This reflects a broader concern that assigning decision-making authority to AI could undermine trust and blur accountability in already ambiguous support roles.

Some participants viewed AI as a tool to reduce the emotional and cognitive load of peer support, suggesting it could be used to draft editable message templates, offer phrasing suggestions, or summarise conversations to support continuity.
However, participants stressed that use should remain flexible and discretionary, valuing AI when it can help structure responses without dictating content or tone, thereby preserving the peer supporter's agency and responsibility.

Participants also raised concerns about AI's ability to provide emotionally attuned support, particularly in culturally specific or linguistically nuanced contexts (\calmsea{}, \flowymoon{}).
Several noted that existing systems struggled with colloquial language or misinterpreted emotional cues, such as by \dquote{trick[ing] the human into thinking it's empathising [...] but it's just sympathy} (\silvercove{}).
Others questioned whether AI could meaningfully build rapport, especially when trust and shared experience were central to support interactions.

Finally, participants observed that many limitations attributed to AI were not unique to automation but reflected broader challenges of text-based and digitally mediated support.
Difficulties in conveying empathy, detecting distress, and sustaining presence were seen as structural constraints of mediated communication.
These reflections suggest that AI systems should be designed not as stand-alone solutions, but as components within responsibility-aware support infrastructures that acknowledge the limits of mediation itself.
\section{Discussion}
\subsection{Reframing Peer Support}
Our findings suggest that definitional plurality is not just semantic but redistributes responsibility and emotional risk.
Where authority is ambiguous, labour accumulates at the individual level, with emotional containment, escalation, and self-care frequently absorbed by supporters rather than distributed across organisational infrastructures.
AI systems that embed fixed assumptions risk stabilising one interpretation of peer support while obscuring this redistribution.

% Our findings show that peer support, as practised, is not a standardised intervention but a relational practice sustained through individual judgement, emotional labour, and informal boundary-setting.
% In the Singapore context, entry pathways were heterogeneous and often grounded in self-identification or lived experience rather than formalised role definitions~\cite{repperReviewLiteraturePeer2011, solomonPeerSupportPeer2004}.
% While this flexibility enabled empathy and trust, it also blurred expectations around scope, authority, and accountability.
% Consequently, responsibility for emotional containment, escalation, and self-care was frequently absorbed by individual supporters rather than distributed across organisational or technical infrastructures.

Prior work documented distinctions within peer identities (e.g., affinity-based peerhood and condition-specific lived-experience roles)~\cite{laraPeerMixedmethodsStudy2026} and divergent conceptions between peer supporters and mental health experts in terms of what peer support is~\cite{simThisReallyHuman2025}.
In such contexts, experts prioritised clinical safety markers and structured validation, whereas peer supporters emphasised relational resonance and authenticity, and AI may implicitly privilege clinical safety markers over relational resonance.
% Participants evaluated AI not primarily in terms of technical capability, but through concerns about responsibility, risk, and labour.
% Aligned with previous work~\cite{simThisReallyHuman2025, wangPracticeOnlinePeer2025}, AI was welcomed when it supported judgement by reducing cognitive load or scaffolding phrasing, but resisted when framed as a replacement for human care or as an autonomous decision-maker.
Participants evaluated AI primarily through concerns about responsibility, risk, and labour, welcoming it when it supported judgement (e.g., reducing cognitive load or scaffolding phrasing) but resisting it as a replacement for human care or an autonomous decision-maker~\cite{simThisReallyHuman2025, wangPracticeOnlinePeer2025}.
These responses challenge AI-supported scaling-oriented narratives and point instead to responsibility-aware systems that make accountability explicit and negotiable.
These findings also raise safety considerations.
In emotionally volatile and ambiguous interactions, risks include misinterpretation of distress, inappropriate escalation, and over-reliance on AI suggestions~\cite{simThisReallyHuman2025}. Responsibility for detecting and correcting such failures remains with peer supporters, creating a misalignment where AI shapes interactions without accountability.
Designing for safety, therefore, requires making responsibility, uncertainty, and escalation pathways explicit within the system.

Finally, peer support is sustained not by individual empathy alone, but by the infrastructures surrounding it.
Where organisations provided supervision, capacity signalling, or clear escalation pathways, participants reported greater confidence and a sense of sustainability.
In their absence, emotional strain intensified, contributing to disengagement.
Technologies for peer support should therefore be embedded within broader care ecosystems rather than positioned as stand-alone conversational agents.
Attending to organisational context, training ecosystems, and the redistribution of labour is essential if technologies are to support peer support practice.

% Taken together, this work reframes peer support technologies as sociotechnical systems that mediate responsibility rather than neutral tools that simply extend reach.
% Foregrounding peer supporters' lived experiences reveals how definitional ambiguity, emotional labour, and boundary-setting are actively negotiated in practice.
% Future work is needed to examine how responsibility is distributed across supporters, organisations, recipients, and technologies over time, particularly as peer support becomes increasingly digitised.
% Such work can inform designs that make responsibility visible, shared, and adjustable, rather than implicitly individualised.

% ---------
\subsection{Design Futures for an AI-Supported Peer Support Ecosystem}
Building on our findings, we outline several design futures for an AI-supported peer support ecosystem that centres not only on the care recipient, but also on the peer supporter as an emotionally and cognitively burdened actor within the system.
Building on our findings, we conceptualise an AI-supported peer support ecosystem as comprising three interrelated system roles:
(1) supporter-facing assistance tools,  
(2) tools that scaffold peer support interactions, and
(3) organisational support infrastructures.

At the level of supporter-facing assistance tools, AI systems can support the emotional well-being of peer supporters.
% Participants described moments of cumulative strain, including repeatedly rereading distressing messages to respond appropriately.
% Rather than positioning AI solely as a tool for drafting responses to clients, future systems could unobtrusively monitor indicators of strain or distress in online contexts, such as prolonged pauses between messages or shifts in linguistic tone, and provide colour-coded or other subtle signals that encourage breaks before burnout occurs.
% In addition, AI could function as a filter during emotionally intense exchanges, generating editable draft templates or phrasing suggestions that reduce the cognitive load of soothing emotions while preserving the supporter’s authorship.
% Where escalation is required, AI could assist with summarising long conversations, supporting structured handoffs to supervisors or therapists, and bridging asynchronous gaps.
% However, such features must be designed with strong privacy safeguards and careful attention to preserving emotional nuance, avoiding reductive or flattened representations of complex experiences.
Beyond drafting tools, systems might detect and track indicators of strain in online exchanges, scaffold emotionally demanding phrasing, facilitate structured handoffs during escalation and surface lightweight prompts suggesting breaks or supervisor check-ins.
Such designs should reduce cognitive load while preserving authorship and emotional nuance, supporting relational labour rather than automating it.
These considerations could potentially also extend beyond text-based settings to multimodal interactions, where strain and meaning may be conveyed through voice, timing, or in-person cues, requiring AI systems to adapt their support (e.g., prompting reflection during calls, summarising across modalities, or supporting transitions between synchronous and asynchronous care).

Tools that scaffold peer support interactions could include editable response drafts, tone-adjustment suggestions, and partial explanations that require the supporter to revise before sending. 
For such tools, maintaining peer supporter agency and not diminishing the human touch (which defines peer support) remains critical.
Interaction designs can incorporate cognitive forcing mechanisms, such as partial explanations accompanying phrasing suggestions, to prompt critical engagement without encouraging blind acceptance~\cite{dejongCognitiveForcingBetter2025}.
Such approaches could help peer supporters evaluate AI outputs more reflectively, balancing support with human discretion.

At the organisational support infrastructure level, AI can provide safeguards to protect peer supporters in online settings.
Systems could detect harassment, boundary violations, or emotionally harmful content directed at supporters, flagging risks and suggesting protective actions, particularly where organisational oversight is uneven~\cite{simSaidThingsNeeded2025}.

Design must also attend to cultural context and lived experience.
Prior work shows that LLM-mediated mental health support often reflects culturally specific assumptions, which may not align with local norms, relational dynamics, or expressions of distress~\cite{songTypingCureExperiences2025, iftikharTherapyNLPTask2024}.
Our participants similarly noted gaps in AI's understanding of colloquial language in the Singaporean context, which may lead to inauthentic responses.
These mismatches shift interpretive burden onto peer supporters, increasing cognitive and emotional labour without redistributing responsibility.

Well-designed systems could instead support interpretation of indirect or colloquial expressions (e.g., flagging ambiguity and suggesting meanings).
Cultural differences in stigma and help-seeking further require differentiated interactional strategies (e.g., interpreting colloquial expressions, adapting tone, or recognising culturally specific help-seeking norms) across demographic groups~\cite{subramaniamStigmaPeopleMental2017, yuanAttitudesMentalIllness2016}.
Future systems should be grounded in locally embedded peer support narratives, potentially for generating suggestions~\cite{hsuHelpingHelperSupporting2025} and simulating clients for training~\cite{yangConsistentClientSimulation2025a}.

At the same time, AI should engage lived experience as both a strength and a boundary.
Systems can scaffold reflection and articulation without displacing ownership for those with lived experience, and support empathetic phrasing while remaining transparent about its affective limitations for those without.
Hybrid interface designs that periodically surface human presence or enable visible handoffs between AI and humans may further sustain trust.

Taken together, these design directions highlight that responsibility in peer support should not be implicitly absorbed by individual supporters or delegated wholesale to AI systems.
Instead, responsibility can be more equitably distributed across an ecosystem of interacting components that support, constrain, and coordinate care practices.

% \subsubsection{Lived Experiences}
% 1. how can AI better support those with lived experiences to 'own' their journey?
% 2. how can AI support those without lived experiences to sound empathetic and not sympathetic (if there is past work saying that humans may also be just offering sympathy instead of empathy especially for novice supporters?)
% 3. PS6 said that AI might pretend to be empathetic but it's just sympathy => can explore how AI can be helpful without pretending to have feelings

% \subsubsection{Others}
% 1. PS17 suggested a "human face behind the AI once in a while" to maintain trust. => can explore this specific hybrid interface design (where AI periodically "hands off" or reveals a human element), having AI presence visible to human
\section{Conclusion}
Peer support is increasingly positioned as a scalable response to gaps in mental health care, yet our findings show that it is shaped by heterogeneous definitions, informal entry pathways, and uneven distributions of responsibility.
Drawing on peer supporters' experiences, we highlight how emotional labour, boundary-setting, and self-care are often individualised in the absence of consistent organisational or technical scaffolding.
Our findings suggest that responsibility, rather than empathy simulation or automation efficiency, is the central design variable in AI-supported peer support. 
Where organisational infrastructures are uneven, technologies risk amplifying hidden burdens under the guise of scale. 
We therefore propose an ecosystem in which AI supports boundary-setting, safeguards supporters, preserves human discretion, and embeds cultural context.
We posit that designing for peer support requires aligning technological intervention with the relational and infrastructural realities of care.

\begin{acks}
This research is supported by the Ministry of Education, Singapore, under its SUTD Kickstarter Initiative (SKI 2021\_04\_08).
Any opinions, findings and conclusions or recommendations expressed in this material are those of the author(s) and do not reflect the views of the Ministry of Education, Singapore.
We would like to thank our reviewers for their thoughtful feedback and constructive suggestions, which helped strengthen this work.
We are also deeply grateful to our participants for sharing their time, experiences, and candid reflections with us.
\end{acks}

\bibliographystyle{ACM-Reference-Format} 
\bibliography{main}

\onecolumn
\appendix
\section{Participant Demographics}
\label{appendix:demographics}
Table~\ref{tab:participant_summary} provides an overview of the participants' demographic backgrounds and peer support experience. 

\begin{table}[ht]
\centering
\small
\caption{Demographic and Peer Support Experience Details}
    \begin{tabular}{@{}p{0.7cm}p{0.9cm}p{0.9cm}p{1.5cm}p{1.3cm}p{2cm}p{3.5cm}p{4cm}@{}}
    \toprule
    \textbf{ID} & \textbf{Age} & \textbf{Gender} & \textbf{Education} & \textbf{Ethnicity} & \textbf{Employment} & \textbf{Psychology Background?} & \textbf{Experience (Type \& Duration)} \\
    \midrule
    \earnprom{} & 25--34 & Female & Bachelor's & Chinese & Full-time & Yes & Online (\textasciitilde1 year) \\
    \bluemoon{} & 45--54 & Female & Bachelor's & Chinese & Full-time & No & Both (\textasciitilde5--10 years) \\
    \warmsun{} & 35--44 & Male & Bachelor's & Chinese & Full-time & No & Offline (\textasciitilde6 years) \\
    \jaderiver{} & 18--24 & Female & Diploma & Chinese & Student & Yes & Both (\textasciitilde4 years) \\
    \windsong{} & 25--34 & Male & Diploma & Chinese & Part-time & No & Both (\textasciitilde1--2 years) \\
    \silvercove{} & 18--24 & Female & Diploma & Malay & Student & No & Both (\textasciitilde2--3 years) \\
    \strongarm{} & 18--24 & Male & Bachelor's & Indian & Student & No & Offline (\textasciitilde6 years) \\
    \clearsky{} & 25--34 & Female & Bachelor's & Chinese & Full-time & Yes & Both (\textasciitilde4 years) \\
    \calmsea{} & 18--24 & Female & Bachelor's & Chinese & Part-time & No & Online (\textasciitilde0.5 years) \\
    \ironpetal{} & 25--34 & Male & Master's & Indian & Full-time & No & Both (\textasciitilde3.5 years) \\
    \echoblaze{} & 18--24 & Female & Bachelor's & Others & Contract & Yes & Both (\textasciitilde3.5 years) \\
    \highpeak{} & 25--34 & Female & Bachelor's & Chinese & Full-time & No & Offline (\textasciitilde3 years) \\
    \shadowdancer{} & 25--34 & Female & Bachelor's & Chinese & Full-time & No & Both (>10 years) \\
    \redbird{} & 25--34 & Female & Bachelor's & Chinese & Part-time & No & Offline (\textasciitilde2--3 years) \\
    \duskytide{} & 25--34 & Female & Bachelor's & Chinese & Self-employed & No & Online (\textasciitilde6 years) \\
    \brightstar{} & 25--34 & Male & Bachelor's & Chinese & Full-time & No & Both (\textasciitilde2--3 years) \\
    \crystalstream{} & 25--34 & Female & Master's & Chinese & Full-time & No & Both (\textasciitilde3--4 years) \\
    \bluepea{} & 18--24 & Female & Bachelor's & Chinese & Student & Yes & Both (\textasciitilde6 years) \\
    \flowymoon{} & 45--54 & Female & Bachelor's & Chinese & Unemployed & No & Offline (\textasciitilde10 years) \\
    \butterflywave{} & 25--34 & Male & Bachelor's & Chinese & Full-time & No & Offline (\textasciitilde1 year) \\
    \bottomrule
    \end{tabular}
\Description{Summary table of 20 peer supporters showing demographics and experience. Participants range from ages 18 to 54, with a majority aged 25-34. Most are female (14 of 20) and predominantly Chinese, with some Malay, Indian, and other ethnicities represented. Education levels include diplomas, bachelor's, and a few master's degrees. Most participants are employed full-time, with some students, part-time workers, and one unemployed participant. A minority have formal psychology backgrounds. Peer support experience varies from approximately 0.5 years to over 10 years, spanning online, offline, and hybrid settings, with most participants engaged in both online and offline support.}
\label{tab:participant_summary}
\end{table}

\section{Semi-Structured Interview Guide}
\label{appendix:interview-guide}
The questions below were used to explore participants' backgrounds, volunteering experiences, reflections on peer support, and challenges.

\paragraph{\textbf{General}}
\begin{enumerate}
\item Do you work in the mental health industry? If not, what field do you work in?
\item How do you think societal attitudes towards mental health have changed over recent years?
\end{enumerate}

\paragraph{\textbf{Volunteering Experience}}
\begin{enumerate}
\item How long have you been volunteering with [organisation]?
\item Do you volunteer with other organisations too?
\item What does your role as a volunteer usually entail?
\item What motivated you to start volunteering in the mental health field?
\item What kind of training did you receive before starting your volunteer work?
\end{enumerate}

\paragraph{\textbf{Peer Support Experience}}
\begin{enumerate}
\item What are some common topics you encounter when providing peer support?
\item Can you describe a particularly rewarding experience you have had while volunteering?
\item Could you share your experiences chatting with clients in physical peer support settings?
\item Could you share your experiences chatting with clients in virtual peer support settings (i.e., through video calls or phone calls, chat-based platforms etc)?
\item What challenges have you faced while talking to clients physically?
\item What challenges have you faced while talking to clients virtually?
\item How do you handle challenging situations or crises while volunteering?
\item What is one thing you enjoy about providing peer support, and why?
\item What is one thing you would like to change about providing peer support, and why?
\item Can you share any feedback from individuals you have supported?
\end{enumerate}

\paragraph{\textbf{Self-Care and Support}}
\begin{enumerate}
\item What kind of support do you receive from [organisation] while volunteering?
\item How do you manage self-care while volunteering in a mentally demanding role?
\end{enumerate}

\paragraph{\textbf{Additional Reflections}}
\begin{enumerate}
\item Are there any other challenges you face as a peer support volunteer?
\end{enumerate}

\end{document}